\documentclass[prl,twocolumn,unsortedaddress,superscriptaddress]{revtex4-1}
\usepackage{graphicx}
\usepackage{epsfig,color}
\usepackage{amsmath,amssymb,eufrak}
\usepackage{amsthm}
\usepackage{enumerate}
\usepackage[colorlinks,urlcolor=orange,citecolor=orange,linkcolor=orange]{hyperref} 

\newcommand{\be}{\begin{equation}}
\newcommand{\ee}{\end{equation}}

\newcommand{\ket}[1]{|{#1}\rangle}

\newcommand{\+}{\! + \!}
\newcommand{\equal}{\! = \!}

\newcommand{\Q}[1]{Q(#1)}
\newcommand{\Qd}[1]{Q^\dag \! (#1)}
\renewcommand{\P}[1]{P(#1)}
\newcommand{\Pd}[1]{P^\dag(#1)}

\definecolor{orange}{rgb}{1,0.5,0}

\begin{document}

\title{Digital quantum simulation of $\mathbb{Z}_2$ lattice gauge theories with dynamical fermionic matter}

\date{\today}

\author{Erez Zohar}
\address{Max-Planck-Institut f\"ur Quantenoptik, Hans-Kopfermann-Stra\ss e 1, 85748 Garching, Germany.}

\author{Alessandro Farace}
\address{Max-Planck-Institut f\"ur Quantenoptik, Hans-Kopfermann-Stra\ss e 1, 85748 Garching, Germany.}

\author{Benni Reznik}
\address{School of Physics and Astronomy, Raymond and Beverly Sackler
Faculty of Exact Sciences, Tel Aviv University, Tel-Aviv 69978, Israel.}

\author{J. Ignacio Cirac}
\address{Max-Planck-Institut f\"ur Quantenoptik, Hans-Kopfermann-Stra\ss e 1, 85748 Garching, Germany.}

\begin{abstract}
We propose a scheme for digital quantum simulation of lattice gauge theories with dynamical fermions. Using a layered optical lattice with ancilla atoms that can move and interact with the other atoms (simulating the physical degrees of freedom), we obtain a stroboscopic dynamics which yields the four-body plaquette interactions, arising in models with $2+1$ and higher dimensions, without the use of perturbation theory. As an example we show how to simulate a $\mathbb{Z}_2$ model in $2+1$ dimensions.
\end{abstract}

\maketitle

Lattice gauge theories are formulations of gauge theories over discretized spacetime \cite{Wilson} or space \cite{KogutSusskind}, that allow either regularization or numerical calculations for continuous gauge theories. They are of particular interest for Quantum Chromodynamics (QCD) \cite{Kogut1983}, the theory of strong interactions, which due to its running coupling \cite{Gross1973} becomes highly nonperturbative at low energies. They have been extremely successful, for example, in calculations for the hadronic (QCD) spectrum \cite{FLAG2013}. However, lattice QCD still faces some computational issues, stemming from the statistical Monte Carlo calculations in Euclidean spacetime on which it is based: first, the computationally hard sign problem \cite{Troyer2005} for a finite fermionic chemical potential;  second, the inability to simulate real-time dynamics when a Euclidean spacetime is used. Clearly these problems have to be overcome in order to reach the ultimate goal of mapping the phases of QCD \cite{Kogut2004,Fukushima2011}, but they also hinder  intermediate steps such as the observation of real-time flux string breaking, which would be a direct manifestation of quark confinement \cite{Wilson,Polyakov1977,Greensite2003}.

Over the last years, new approaches for dealing with lattice gauge theories have emerged from the communities of quantum information and quantum optics. One of them is quantum simulation \cite{Feynman1982}, in which lattice gauge theories are mapped to systems that can be accurately controlled in the lab, such as cold atoms in optical lattices \cite{Bloch2008,Lewenstein2012}, trapped ions \cite{Lanyon2011,Schneider2012}, superconducting qubits \cite{Romero2016} or Rydberg atoms \cite{rydberg}.
This can be done in either an analog way, in which both the degrees of freedom and dynamics are mapped to those of a quantum simulator
\cite{Zohar2015a,Wiese2013,Zohar2011,Zohar2012,Banerjee2012,Zohar2013,NA,Topological,Rishon2012,AngMom,ZollerIons,SQC,Dissipation,Kosior2014,Marcos2014,Wiese2014,Pepe2015}, or digitally \cite{Jane2003a}, where trotterized dynamics is obtained by a stroboscopic series of quantum operations \cite{rydberg,Tagliacozzo2013,TagliacozzoNA,Mezzacapo2015}.


Digital quantum simulations have been very successful recently for the study of several many body models (e.g. \cite{Lanyon2011,Salathe2015,Barends2015,Barends2016}), including a recent realization, using trapped ions,  of the lattice Schwinger model ($1+1$ QED) \cite{Martinez2016}.

Lattice gauge theories in $2+1$ dimensions and above set a challenge for quantum simulation: they involve a four-body interaction term, the magnetic plaquette interaction, whose implementation is nontrivial. In analog simulations, such interactions are typically obtained effectively,
by the use of a fourth-order perturbative Hamiltonian construction \cite{AngMom}, but this automatically makes them weak and sets an experimental bottleneck.

In this work we introduce a different scenario for quantum simulation that circumvents this bottleneck. We discuss a way to construct stroboscopically the dynamics of a lattice gauge theory, coupled to dynamical fermionic matter, and present, as a first example, a possible implementation of the simple case of
$\mathbb{Z}_2$ using ultracold atoms (which allow us to work naturally with fermions as required by lattice gauge theories with more than $2+1d$). The atoms will be trapped in a layered structure, and will include, on top of atomic species representing the matter and gauge field degrees of freedom, some ancillary atoms that will be moved between the different layers, as proposed in \cite{Aguado2008}. This will allow to obtain stroboscopically the desired interactions without the use of perturbative arguments, while avoiding any undesired interactions, making the plaquette interactions stronger than in the previous analog proposals, hence suggesting a path towards quantum simulations in 2+1 dimensions and more. Other models will be discussed in \cite{StatSim}.

$\mathbb{Z}_N$ lattice gauge theories \cite{Horn1979} are relevant for two reasons: first, they can be seen as a truncation method for compact QED  which is restored in the $N\rightarrow\infty$ limit \cite{Horn1979}; second, $\mathbb{Z}_N$ is the center of $SU(N)$, and as such it plays a crucial role in confinement \cite{Hooft1978,Greensite2003}.
$\mathbb{Z}_2$ is the simplest case of such a model. For the second reason, $\mathbb{Z}_2$ theories without dynamical matter have been studied, and were shown to have a confining phase for static charges \cite{Horn1979}. Quantum simulations of the type we propose might be able to extend the confinement study to dynamical charges as well.
A pure gauge $\mathbb{Z}_2$ (without dynamical matter) digital quantum simulation was proposed in \cite{Tagliacozzo2013}. A similar Hilbert space truncation was in the $1+1$ dimensional link model \cite{Horn1981,Wiese1997} quantum simulation \cite{Banerjee2012}. Another relevant previous work deals with the quantum simulation of quantum double models \cite{Aguado2008}.

\begin{figure}[t!]
\begin{center}
	\includegraphics[width=0.4\textwidth]{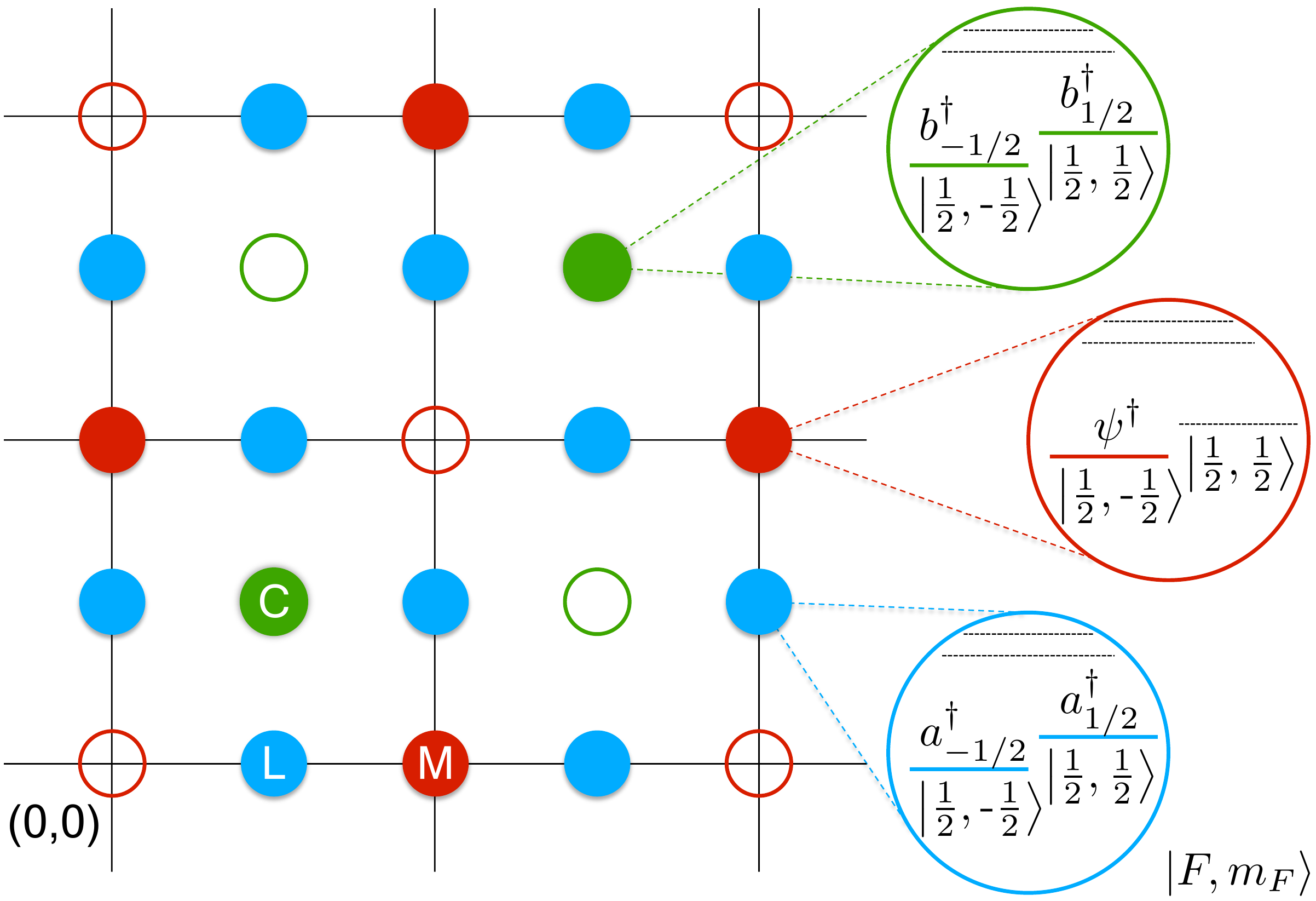}\vspace{10pt}\\
	\includegraphics[width=0.3\textwidth]{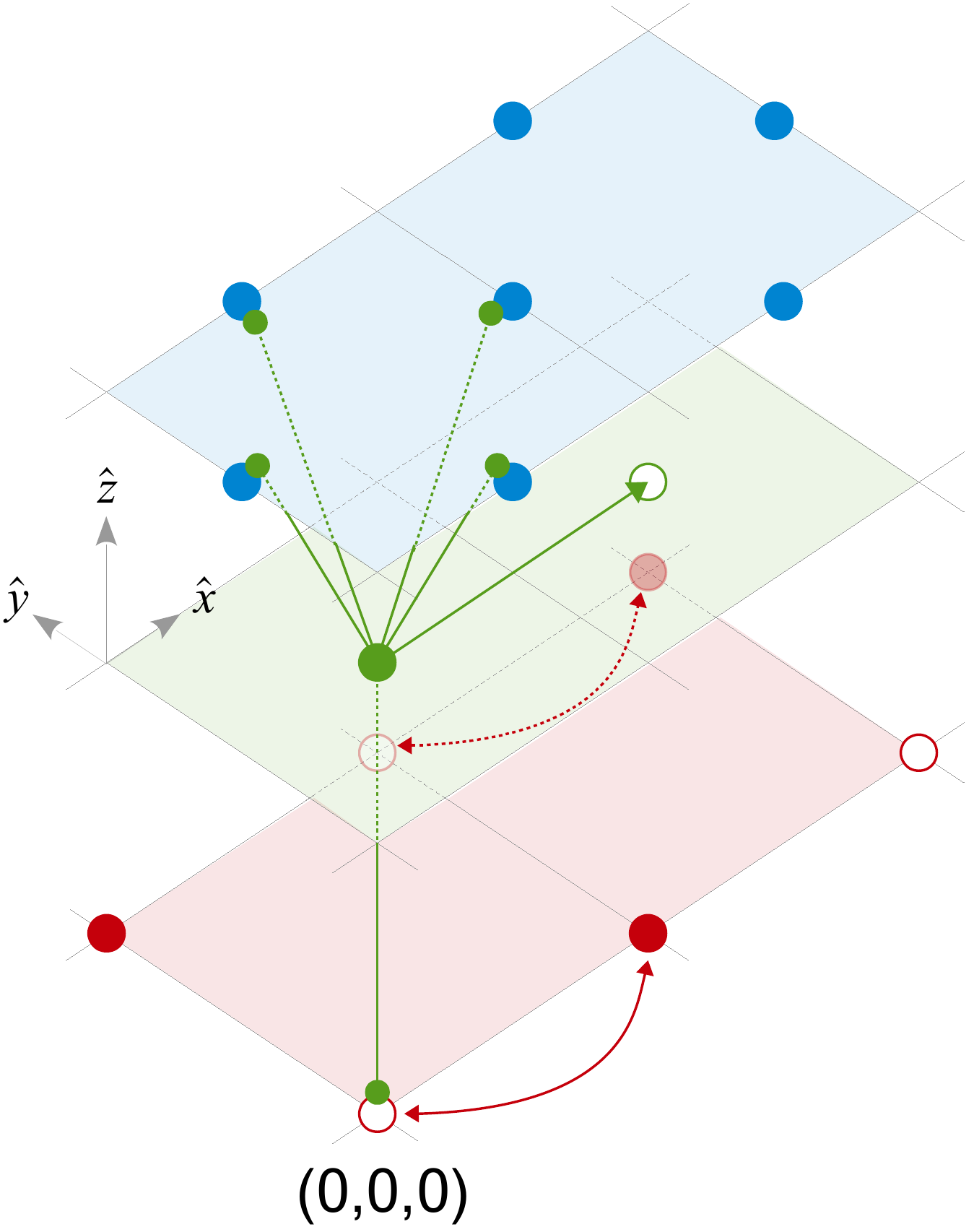}\includegraphics[width=0.12\textwidth]{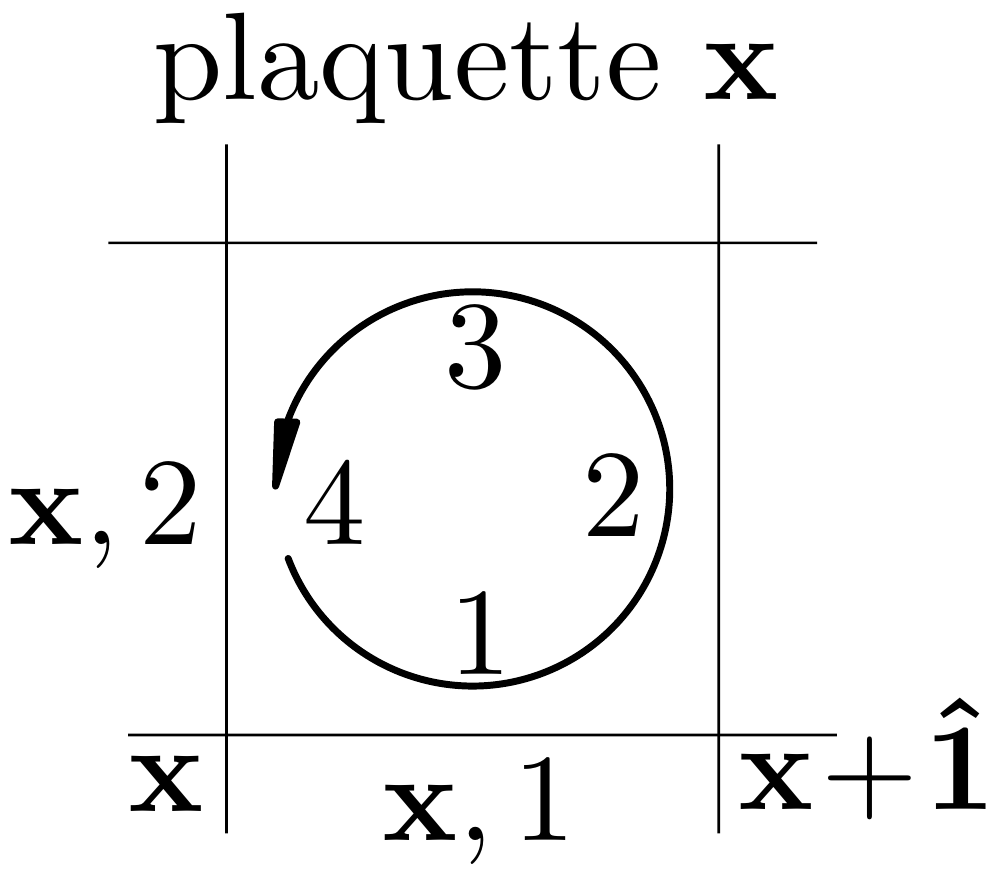}
	\begin{minipage}[t]{0.5\textwidth}
	\caption{(Top) Initial position of the different atoms with respect to the simulated lattice. Full/empty circles represent initially occupied/empty potential minima. The matter fermions (red) are trapped in correspondence with the odd sites (half-filling). The link atoms (blue) are trapped in correspondence with the middle of the links. The auxiliary atoms (green) are trapped in correspondence with the middle of even plaquettes. The internal structure of each atomic species is shown on the right, with the relevant atomic levels highlighted.\\(Bottom left) Different atomic species reside on different vertical layers. Green straight lines show how the auxiliary atoms have to move in order to realise interactions with the link atoms and the fermions, or to enter odd plaquettes. Red arrows show selective tunneling of fermions across even horizontal links.\\(Bottom right) Labeling convention for vertices and links, shown for one plaquette.}
	\label{FigLattice}
	\end{minipage}
\end{center}	
\end{figure}

\emph{$\mathbb{Z}_N$ lattice gauge theory.} Consider two unitary operators $P$ and $Q$, defined in an $N$-dimensional Hilbert space and satisfying the relations $P^N=Q^N=1$, $PQP^{\dagger}=e^{i\delta}Q$ with $\delta=2\pi/N$. If we define a basis of $P$ eigenstates,
$P\left|m\right\rangle = e^{i\delta m}\left|m\right\rangle$, $Q$ is a \emph{cyclic} raising operator, $Q\left|m\right\rangle=\left|m+1\right\rangle$.This $\mathbb{Z}_N$ algebra is the basis of a $\mathbb{Z}_N$ lattice gauge theory. Consider a spatial lattice, e.g. in two dimensions, and place such a Hilbert space on each of its links, which we label by the coordinate-direction pair $\mathbf{x},k$ (see Fig.~\ref{FigLattice}).
The dynamics of this system is described by the Hamiltonian ${ H_{EB}=H_E+H_B }$ \cite{Horn1979}, where
$H_E = \lambda_E\underset{\mathbf{x},k}{\sum}\left(1-P\left(\mathbf{x},k\right)-P^{\dagger}\left(\mathbf{x},k\right)\right)$
is called the electric Hamiltonian, and
${ H_B \equal \lambda_B\underset{\mathbf{x}}{\sum}\left( \Q{\mathbf{x},1} \Q{\mathbf{x \+ \hat{1}},2} \Qd{\mathbf{x \+ \hat{2}},1} \Qd{\mathbf{x},2} \+ H.c.\right) }$
is called the magnetic Hamiltonian. A proper choice of $\lambda_{E,B}$ reproduces the $U(1)$ Kogut-Susskind Hamiltonian \cite{Kogutlattice} for $N\rightarrow\infty$
\cite{Horn1979}.

 The gauge field may also be coupled to fermions. These reside on the lattice's vertices, labeled by the coordinates $\bf{x}$, and are created by the fermionic operators $\psi^{\dagger}(\mathbf{x})$. Their dynamics, including the coupling to the gauge field, is given by ${ H_F = H_M+H_{GM} }$, where
${ H_M=M\underset{\mathbf{x}}{\sum}\left(-1\right)^{x_1+x_2}\psi^{\dagger}\left(\mathbf{x}\right)\psi \left(\mathbf{x}\right)}$
is the local mass term, and
${ H_{GM}=\lambda_{GM}\underset{\mathbf{x},k}{\sum}\left(\psi^{\dagger}\left(\mathbf{x}\right) \Q{\mathbf{x},k} \psi (\mathbf{x+\hat{k}} )+H.c\right) }$
is the gauge-matter coupling.
The fermions are staggered \cite{Susskind1977,Zohar2015}, and carry the charge
${ q\left(\mathbf{x}\right)=\psi^{\dagger}\left(\mathbf{x}\right)\psi\left(\mathbf{x}\right)
-\left(1-\left(-1\right)^{x_1+x_2}\right)/2 }$.
The total Hamiltonian, $H=H_{EB}+H_F$ is invariant under local gauge transformations, i.e. $H=G\left(\mathbf{x}\right)H G^{\dagger}\left(\mathbf{x}\right)$ for every $\mathbf{x}$, with
\begin{equation}
G\! \left(\mathbf{x}\right) \equal \P{\mathbf{x},1} \P{\mathbf{x},2} \Pd{\mathbf{x \!-\! \hat{1}},1} \Pd{\mathbf{x \!-\! \hat{2}},2} e^{- i \delta q\left(\mathbf{x}\right)}
\end{equation}
(analogous to the Gauss law for continuous groups).
In the $\mathbb{Z}_2$ case, the link becomes a two-level system and we can identify $P=\sigma_z=P^\dag$ and $Q=\sigma_x=Q^\dag$ (both unitary and Hermitian).
Then, the Hamiltonian terms become simpler,
 ${ H_E = \lambda_E\underset{\mathbf{x},k}{\sum}\left(1-2\sigma_z\left(\mathbf{x},k\right)\right) }$ and
 ${ H_B=2\lambda_B\underset{\mathbf{x}}{\sum}\sigma_x\!\left(\mathbf{x},1\right)\sigma_x\!\left(\mathbf{x\+\hat{1}},2\right)\sigma_x\!\left(\mathbf{x\+\hat{2}},1\right)\sigma_x\!\left(\mathbf{x},2\right) }$.
Although we discussed above the $2+1d$ case, it can be generalized to higher dimensions.

\emph{Digitization.} Under the action of the total Hamiltonian $H$, the simulated model evolves with the unitary operator ${ U(t) = e^{-i H t} }$. However, it is hard to map the total Hamiltonian to cold atomic terms, and thus a digital simulation can be helpful:  If a Hamiltonian $H = \sum_j H_j$ decomposes into terms $H_j$, then by Trotter formula~\cite{Trotter1959a,Suzuki1985a,Bhatia1996}  ${ e^{-i H t} \!=\! \lim_{M\rightarrow \infty} \!\big( \mathbf{\Pi}_j \; e^{-i H_j t/M} \big)^M \!\equiv\! \lim_{M\rightarrow \infty} \!\big( W(\frac{t}{M}) \big)^M }$
 we can approximate the total evolution with a series of short evolutions $W_j = e^{-i H_j \tau}$, where $§\tau=t/M$  \cite{LLoyd1996a,Jane2003a}.
As we will show, we can implement the parts of $H$ separately, and approximate the desired evolution with a suitable sequence of the unitary operators $W_j$.

%

\emph{Obtaining effective plaquette interactions.}
The interacting parts of $H$, i.e. $H_B$ and $H_{GM}$, are generated in our scheme using intermediate interactions with some ancillary degrees of freedom that we add in the middle of the plaquettes. These ancillas (labeled by a tilde superscript in the following) have a ``copy'' of the link Hilbert space, i.e. we can define the $\tilde{P},\tilde{Q}$ operators and the \mbox{$\tilde{P}$-eigenstates} $\left|\tilde{m}\right\rangle$. Their role can be conveniently described by an object called ``stator" \cite{Reznik2002,ZoharStators}, which is a mixture of an operator in one Hilbert space and a state in another one, and results from the interaction of a link and an ancilla. Let us focus on a single plaquette $\mathbf{x}$, and label by $1$-$4$ its links, counter-clockwise, starting from $\left(\mathbf{x},1\right)$ and ending with $\left(\mathbf{x},2\right)$ (see Fig.~\ref{FigLattice}). We define the following unitary operators, acting on the ancilla and the link $i$:
\begin{equation}
\mathcal{U}_i = \underset{m}{\sum} Q_i^m \otimes \left|\tilde{m}\right\rangle\left\langle\tilde{m}\right| .
\label{Ueq}
\end{equation}
Suppose the ancilla is initially prepared in the state $\left|\tilde{in}\right\rangle=\frac{1}{\sqrt{N}}\underset{m}{\sum}\left|\tilde{m}\right\rangle$. Then, we define a stator for the $i$ link as $S_i=\mathcal{U}_i\left|\tilde{in}\right\rangle$. Note that this object satisfies the \emph{eigenoperator relation} ${ \tilde{Q}S_i = S_iQ^{\dagger}_i }$, that is  ${ (\tilde{Q} \otimes \mathbb{I}_i ) \mathcal{U}_i (\ket{\tilde{in}} \otimes \ket{\psi_i}) = \mathcal{U}_i (\ket{\tilde{in}} \otimes Q_i^\dag \ket{\psi_i}) }$ for any state $\ket{\psi_i}$ of the link. This allows us to obtain effectively the plaquette dynamics of $H_B$, by acting with a sequence of $\mathcal{U}_i$ operators and then performing a local operation on the ancilla. Define $Q_{\square}=Q_1Q_2Q^{\dagger}_3 Q^{\dagger}_4$, and
\begin{equation}
S_{\square} \equiv \mathcal{U}_{\square} \left|\tilde{in}\right\rangle \equiv \mathcal{U}_1 \mathcal{U}_2 \mathcal{U}^{\dagger}_3 \mathcal{U}^{\dagger}_4 \left|\tilde{in}\right\rangle \equal \frac{1}{\sqrt{N}}\underset{m}{\sum}Q^m_{\square}\otimes\left|\tilde{m}\right\rangle .
\label{Seq}
\end{equation}
This ``plaquette stator" satisfies $\tilde{Q}S_{\square} = S_{\square}Q_{\square}^{\dagger}$. Then, if we act locally on the ancilla with
$\tilde{V}_B = e^{-2 i\lambda_B \tau \left(\tilde{Q}+\tilde{Q}^{\dagger}\right)}$, we obtain the desired plaquette interaction,
since
\begin{equation}
\tilde{V}_B S_{\square} = S_{\square} e^{-i H_B \tau}
\end{equation}
and thus $\mathcal{U}^{\dagger}_{\square} \tilde{V}_B \mathcal{U}_{\square}\left|\tilde{in}\right\rangle$ will simply give rise to
$\left|\tilde{in}\right\rangle e^{-i H_B\tau}$.
To recap, we first let all the links interact separately with the ancilla prepared in $\ket{\tilde{in}}$ and create $\mathcal{U}_{\square}$, then we act locally for time $\tau$ on the ancilla with $\tilde{H}_B$ and finally let the links interact again to form $\mathcal{U}_{\square}^{\dagger}$. This way we obtain effectively the plaquette dynamics that we wanted, and the ancilla is back in its initial state, completely disentangled from the links and ready for the next step of the digital simulation: pairwise interactions of the ancilla with each of the links around it gave rise to a four body interaction, without any use of perturbation theory. Note that this procedure can be done in parallel for all the plaquettes belonging to the same parity (since every link is associated with two plaquettes).

\emph{Obtaining effective gauge-matter interaction.}
The interaction between the gauge fields and the matter is generated in a similar way. This time, we only need a stator for a single link, $S_i$. Once the stator is ``on", the ancilla interacts with the fermion $\psi$ at one end of the relevant link, to generate the unitary $\mathcal{U}^{\dagger}_W = e^{ -\psi^{\dagger}\psi \log\left(\tilde{Q}^{\dagger}\right)}$, then the fermions at both ends of the link (call the one at the other end $\chi$) are allowed to tunnel for some time $\tau$, giving rise to $\mathcal{U}_{t}=e^{-i \lambda_{GM} \left(\psi^{\dagger}\chi + \chi^\dag \psi \right)\tau}$ and finally we act with $\mathcal{U}_W$. Altogether, $\mathcal{U}_W \mathcal{U}_{t} \mathcal{U}_W^{\dagger} = e^{-i\lambda_{GM}\left(\psi^{\dagger}\tilde{Q}^{\dagger}\chi + H.c.\right)\tau}$ and when we act with it on the stator
\begin{equation}
\mathcal{U}_W \mathcal{U}_{t} \mathcal{U}_W^{\dagger} S_i =
S_i e^{-i\lambda_{GM}\left(\psi^{\dagger}Q_i\chi + H.c.\right)\tau}
\end{equation}
we get the on-link interaction from $H_{GM}$. The stator is then ready for the next task or cancellation with $\mathcal{U}_i^{\dagger}$.

\emph{The simulating system.} We now propose a simulation of $\mathbb{Z}_2$ based on ultracold atoms in optical lattices. Although the $\mathbb{Z}_2$ model is planar, the simulation will take place across several layers, separating the experimental ingredients in a way that avoids undesired interactions while moving the control atoms (ancillas) (see Fig. \ref{FigLattice}).

Layer 1 (lowest): fermionic matter. Fermions, created by the operators $\psi^{\dagger}\left(\mathbf{x}\right)$, can be trapped at the vertices of the simulated lattice. The fermionic optical lattice is half-filled, and the minima are very deep such that fermions cannot tunnel to other vertices. However, it is possible to reduce the energy barrier between neighboring wells in a way that allows tunneling. This can be done separately for four types of links, which we denote by even/odd (according to the parity of the vertex from which they emanate) and horizontal/vertical. Initially, only the odd vertices are full, representing a Dirac sea state \cite{StatSim} We assume that the physical fermions all belong to one hyperfine level (e.g. $\left|F'=1/2,m'_F=1/2\right\rangle$) of a fermionic alkaline atom.

Layer 2 (highest): Gauge fields. These are atoms representing the gauge degrees of freedom, trapped on the links of the lattice. We place exactly one atom per link and make the minima very deep so that they cannot tunnel. To get a 2-dimensional Hilbert space, we need control over two internal levels: these can be the $\left|F=1/2,m_F=\pm 1/2\right\rangle$ levels of a second fermionic alkaline species in its hyperfine ground state
(note that in general one may use other hyperfine multiplets, even bosonic ones, and effectively eliminate some levels to obtain a similar structure.)
 These levels represent the $P$ eigenstates of the link's Hilbert space. The corresponding fermionic creation operators are $a^{\dagger}_m\left(\mathbf{x},k\right)$, giving rise to the second-quantized angular momentum operators $F^{\alpha}\left(\mathbf{x},k\right)=\frac{1}{2} \sigma^{\alpha}\left(\mathbf{x},k\right)=\frac{1}{2}a^{\dagger}_m\left(\mathbf{x},k\right) \sigma^{\alpha}_{mn} a_n\left(\mathbf{x},k\right)$.

Layer 3 (middle): Control atoms. These belong to a third atomic species, and are trapped in the center of the even plaquettes. These atoms are movable in two ways: first, they may be moved to interact with the gauge bosons on the links around them and with the matter fermions on their left-bottom (see Fig. \ref{FigLattice}); second, their "rest position" may be moved to the centers of the odd plaquettes. The control atoms must also be described by a 2-dimensional Hilbert space (as above). Again we can use  two $\left|\tilde{F}=1/2,\tilde{m}_F=\pm 1/2\right\rangle$ levels.
The corresponding  creation operators are $b^{\dagger}_m\left(\mathbf{x}\right)$, giving rise to the second-quantized angular momentum operators ${ \tilde{F}^{\alpha}\left(\mathbf{x}\right)=\frac{1}{2}\tilde{\sigma}^{\alpha}\left(\mathbf{x}\right)=\frac{1}{2}b^{\dagger}_m\left(\mathbf{x}\right) \sigma^{\alpha}_{mn} b_n\left(\mathbf{x}\right) }$.

We assume that the relevant $m_F$ levels are splitted by either a background static magnetic field, or an AC-stark effect, which must produce different energy splittings for the three atomic species. Further details on the trapping potentials may be found  in \cite{StatSim}.

\emph{Experimental realization of the digital sequence.}
The different parts of the Hamiltonian described above are created by the use of either local operations or interactions between the control and the physical atoms. We now introduce a set of local operations and interactions out of which the desired evolution will be built:

\emph{1) Local `laser" terms.} These are local operations, generated, for example, by connecting different atomic levels with Raman lasers. We define
  ${ V_{\mathbf{n}}\left(\phi\right) = e^{-i\phi\underset{\mathbf{x}}{\sum}\mathbf{n}\cdot\sigma\left(\mathbf{x}\right)} }$
  as the result of acting with a Hamiltonian $H_{\mathbf{n}}=\lambda \mathbf{n}\cdot\sigma$ on all the link atoms simultaneously, for a time $t=\phi/\lambda$. Similarly, we can define $\tilde{V}_{\mathbf{n}}\left(\phi\right)=e^{-i\phi\underset{\mathbf{x}}{\sum}\mathbf{n}\cdot\tilde{\sigma}\left(\mathbf{x}\right)}$ for the control atoms.

 \emph{2) $a$-$b$ Scattering terms}, resulting from \mbox{S-wave} collisions of $a$ (link) and $b$ (control) atoms. To generate these, we move the control atoms to a link such that it overlaps with the physical atom and after some time we move it back. This induces a two-body collision, which is described by the Hamiltonian
    ${ H_{ab}=f_0\left(t\right)\left(g_0 \underset{m,n}{\sum}a^{\dagger}_ma_m b^{\dagger}_nb_n + g_1 \mathbf{F}\cdot\tilde{\mathbf{F}}\right) }$
    where $f_0\left(t\right)$ is the result of the time-dependent overlap of the Wannier functions, 
    and $g_{0,1}$ depend on the S-wave scattering lengths $a_{0,1}$ and the reduced mass $\mu$ of the two atoms~\footnote{$g_0 = \pi \left(a_0+3a_1\right) / 2\mu$,
    $g_1 = 2\pi \left(a_1-a_0 \right) / \mu$}.
  Since tunneling is not allowed, $\underset{m}{\sum}a^{\dagger}_ma_m = \underset{m}{\sum} b^{\dagger}_mb_m=1$ and the first term gives rise to a global phase, which can be ignored.
     Moreover the energy splittings for the $a$ and $b$ atoms are different and thus, in a rotating wave approximation, we can consider only the $z$ part of the second term. We eventually obtain a unitary of the form
     $\mathcal{U}_{\text{ab}}\left(\phi\right)=e^{-4i \phi F_z \tilde F_z}=e^{-i \phi \sigma_z \tilde \sigma_z}$
     which, when combined with local unitaries, will give rise to the creators of stators, $\mathcal{U}_i$. Further details are given in the appendix.

    \emph{3) $b$-$\psi$ scattering terms.} These are similar to the above, with a scattering Hamiltonian taking the form
    $H_{b\psi}=f'_0\left(t\right)\left(g'_0 \psi^{\dagger}\psi\underset{m}{\sum}b^{\dagger}_mb_m+g'_1\psi^{\dagger}\psi\tilde{\sigma}_z\right)$.
    Note that the total $b$ number is 1 and ${ \log\tilde{\sigma}_z = i\pi\left(1-\tilde\sigma_z\right)/2 }$. Thus we may write this interaction as
    ${ H_{b\psi}=f'\left(t\right)\left(\left(g'_0+g'_1\right) \psi^{\dagger}\psi +\frac{2ig'_1}{\pi}\psi^{\dagger}\psi\log\tilde{\sigma}_z\right) }$.
    This gives rise to a unitary $\mathcal{U}_{b\psi}\left(\phi\right)=e^{-i\phi'\left(\phi\right)\psi^{\dagger}\psi}e^{\frac{-\phi}{\pi}\psi^{\dagger}\psi\log\tilde{\sigma}_z}$
   which, along with local unitary operations, will result in the $\mathcal{U}_W$ operators required for the gauge-matter interactions (see the appendix for further details.)

With these at hand, we can finally write down expressions for the required operations. The non-interacting ones will take the forms
$W_E = e^{-i H_E\tau} = V^{\dagger}_z\left(2\tau\lambda_E\right)$,
and
$W_M = e^{-i H_M\tau}$
which is achievable by changing the depth of the optical trap for the vertex fermions in a staggered fashion.
The interactions are based on the atomic collisions introduced above, which are used for the construction of the unitary operations $\mathcal{U}_i$, creating a stator on a link, as well as $\hat{\mathcal{U}}$, the unitary responsible for the link interactions. More details are found in the appendix.

\emph{The complete sequence.}
The stroboscopic dynamics will eventually consist of eight pieces: $W_{E},W_{M}$, the local terms; $W_{Be},W_{Bo}$, responsible for the plaquette interactions on even and odd plaquettes respectively; $W_{ev},W_{eh},W_{ov},W_{oh}$, responsible for the gauge-matter interactions on even/odd horizontal/vertical links. A detailed description of the sequence is given in the appendix. Using the commutativity of several of the unitaries in this sequence ($W_{B,e},W_{B,o}$ with anything but $W_E$, $W_M$ with $W_E,W_B$) and defining ${ W_B=W_{B,o}W_{B,e} }$, we obtain the time evolution of a sequence - i.e. a single time step of the digitized evolution:
\begin{equation}
W\left(\tau\right) = W_M W_E W_B W_{oh} W_{ov} W_{eh} W_{ev} .
\end{equation}
All terms in the sequence respect the local gauge symmetry, i.e. each term individually commutes with the $G\left(\mathbf{x}\right)$ operators for every $\mathbf{x}$.
Thus, symmetry is not affected by the digitization, at least in the ideal case.


Several types of errors can however affect the simulation.
 First, intrinsic errors coming from the digitization. The sequence described above is only equal to the desired evolution to first approximation, and errors scaling as $t^2/M$ arise because the various ingredients of the sequence do not commute~\cite{Suzuki1985a,LLoyd1996a,Jane2003a}. However, the error can be made as small as desired by increasing the number of steps $M$, at the cost of a longer simulation time. It is also important to remind that each piece of the sequence, and hence their commutators, is gauge invariant so the evolution of the simulating system ideally respects the desired symmetry. Second, there might be experimental imperfections, i.e. the implementation of the desired local evolutions and interactions can deviate from the expected one. Importantly, such errors can break the symmetry, so care should be taken in minimizing this effect. Errors may either scale proportionally to the infinitesimal time step $\tau$ (in the cases of $\mathcal{U}_t,\tilde{V}_B,W_E,W_M$), or be of order $1$ (independent of $\tau$) in the cases of $\mathcal{U}, \mathcal{U}_W$ etc. The latter type of error is more dangerous since they can accumulate for a large $M$: $M$ will depend on the size of the system, and thus the errors will have to scale inversely with the system size \cite{StatSim}. Among the possible sources of error: atoms can be excited to other internal levels due to the trapping fields; atoms must be moved adiabatically to avoid excitations \cite{Jaksch1999};
    each step must be accurately timed etc. Yet, one has to bear in mind that errors are random, as well as that for quantum simulation purposes (rather than quantum computation) the effect of errors and tradeoffs may be better tuned and less harmful.
      Further discussion of the errors may be found in \cite{StatSim}.

 \emph{Summary.}
 In this paper we presented a way to construct a cold atomic quantum simulator of a $\mathbb{Z}_2$ lattice gauge theory with fermionic matter in $2+1$ dimensions, which relies on the implementation of a stroboscopic dynamics mediated by ancillary degrees of freedom. A convenient description of the system is proposed through the formalism of stators.

 This approach has several advantages. It consists of simple ingredients, aligned in a layered structure. Thus, the gauge, matter and control atoms may be trapped, controlled, manipulated and measured separately, in a clean way.
 The stroboscopic dynamics consists only of gauge invariant steps, and thus the gauge symmetry is in principle unaffected by digitization. The gauge invariant interactions  are achieved by the use of ancillas, allowing to avoid perturbation theory. Furthermore, the implementation does not require sophisticated experimental techniques such as the use of Feshbach resonances and can be extended to other models with both abelian and non-abelian gauge
symmetries.


\emph{Acknowledgements}. The authors would like to thank Tao Shi and Andr\'as Molnar for helpful discussions. EZ acknowledges the support of the Alexander-von-Humboldt foundation.

\section*{Appendix}
In this appendix, we give details on the sequence of operations required for the creation of the stroboscopic time evolution. First we consider the creation, from the building blocks presented in the main text, of the unitary operations $\mathcal{U}_j$, $\mathcal{U}_W$, $\hat{\mathcal{U}}$, responsible to single steps in our sequence. Then we will turn to the construction of the whole sequence.

\begin{widetext}

 \subsection{Creating the unitary operations}

 All operations are based on the control atoms prepared in an initial state
\begin{equation}
  \left|\tilde{in}\right\rangle = \left|\uparrow_x\right\rangle = \frac{1}{\sqrt{2}}\left(\left|0\right\rangle + \left|1\right\rangle\right)
\end{equation}
and, as may be seen below, in the end of each interaction step the control atoms will go back to this initial state.

The stators are generated  (up to a global phase) by the sequence
\begin{equation}
\mathcal{U} = V^{\dagger}_y\left(\frac{\pi}{4}\right) \mathcal{U}_{\text{a}_{j}\text{b}}\left(\frac{\pi}{4}\right) V_y\left(\frac{\pi}{4}\right) V_x\left(\frac{\pi}{4}\right)
\tilde{V}_z\left(\frac{\pi}{4}\right),
\end{equation}
where the local operations $V_k$ and $\tilde{V}_k$ act simultaneously on all the link or on all the control atoms (respectively), while $\mathcal{U}_{\text{a}_{j}\text{b}}$ involves only the atoms sitting on the link $j$ of even/odd plaquettes (this is achieved by moving the control atoms to overlap with atoms on link $j$).  Undoing the stators is simply done with the same sequence, as for $\mathbb{Z}_2$, $\mathcal{U}^{\dagger}=\mathcal{U}$ (see eq.(2)). Note, as well, that if one wishes to create a plaquette stator, some of the steps may be shared by all the links and do not have to be repeated:
\begin{equation}
\mathcal{U}_{\square} = V^{\dagger}_y\left(\frac{\pi}{4}\right) \mathcal{U}_{\text{a}_{1}\text{b}}\left(\frac{\pi}{4}\right) \mathcal{U}_{\text{a}_{2}\text{b}}\left(\frac{\pi}{4}\right) \mathcal{U}_{\text{a}_{3}\text{b}}\left(\frac{\pi}{4}\right) \mathcal{U}_{\text{a}_{4}\text{b}}\left(\frac{\pi}{4}\right)V_y\left(\frac{\pi}{4}\right) V_x\left(\frac{\pi}{4}\right)
\tilde{V}_z\left(\frac{\pi}{4}\right).
\end{equation}

For the interaction with the matter, note that
\begin{equation}
\tilde{V}_y\left(\frac{\pi}{4}\right)\mathcal{U}_{b\psi}\left(\pi\right)\tilde{V}^{\dagger}_y\left(\frac{\pi}{4}\right)=e^{-i \phi'\psi^{\dagger}\psi}\mathcal{U}^{\dagger}_W
=\mathcal{U}^{\dagger}_We^{-i \phi'\psi^{\dagger}\psi},
\end{equation}
and that
\begin{equation}
\tilde{V}^{\dagger}_y\left(\frac{\pi}{4}\right)\mathcal{U}_{b\psi}\left(\pi\right)\tilde{V}_y\left(\frac{\pi}{4}\right)=e^{-i \phi''\psi^{\dagger}\psi}\mathcal{U}_W.
\end{equation}
Then, the link interaction may be realized by the sequence
\begin{equation}
\hat{\mathcal{U}}=\tilde{V}^{\dagger}_y\left(\frac{\pi}{4}\right)
\mathcal{U}_{b\psi}\left(\pi\right)
\tilde{V}_y\left(\frac{\pi}{2}\right)
\mathcal{U}_{t}\left(\lambda_{GM}\tau\right)
\mathcal{U}_{b\psi}\left(\pi\right)
\tilde{V}^{\dagger}_y\left(\frac{\pi}{4}\right)
= e^{-i \phi''\psi^{\dagger}\psi}
\mathcal{U}_W
\mathcal{U}_{t}\left(\lambda_{GM}\tau\right)
\mathcal{U}^{\dagger}_W
e^{-i \phi'\psi^{\dagger}\psi}
\end{equation}
acting on the stator.
The phase factors $e^{-i \phi'\psi^{\dagger}\psi},e^{-i \phi''\psi^{\dagger}\psi}$ have no influence on the physics once considered over the entire lattice,
as explained in \cite{StatSim}.

\clearpage
\end{widetext}

\subsection{The complete experimental sequence}

The stroboscopic sequence of a single time step, equivalent to simulated time $\tau$, is as follows: \\
 1. Start with the controls placed in the centers of the even plaquettes. Interact with the link on the left ($\mathcal{U}_4$), to create a stator $S_4$ for all the even plaquettes. Then perform the sequence $\hat{\mathcal{U}}$ for this link, to realize the gauge matter interactions of even vertial links -
      $W_{ev}$. Then act with $\mathcal{U}_4$ again to undo the stator.\\
 2. Repeat a similar process for the link below ($\mathcal{U}_1$), but without undoing the stator: this will create $W_{eh}$, the $H_{GM}$ evolution for even horizontal links.\\
 3. Interact with the remaining links to create the plaquette's stator (for all the even plaquettes). Then perform the control time evolution $\tilde{V}_B = \tilde{V}_x\left(2\lambda_B\tau\right)$ to obtain $W_{B,e}$ - the unitary responsible for the time evolution of $H_B$ on even plaquettes. Remove the interaction with all the links, except the one on the right.\\
  4. Move the controls to the right until they reach the center of the odd plaquettes. The stator for the odd vertical links is ready, so its interactions may be completed in a similar manner to these of step 1, to obtain $W_{ov}$.\\
  5. Repeat the procedure of step 2 to obtain $W_{oh}$. Again, do not undo the stator for the lower link.\\
  6. Complete the plaquette's stator, act with $\tilde{V}_B$, and obtain $W_{B,o}$. Undo the plaquette.\\
  7. Perform the local unitaries, $W_E$ and $W_M$.\\

This gives the full sequence
\begin{equation}
	W_M W_E W_{B,o} W_{oh} W_{ov} W_{B,e} W_{eh} W_{ev}.
\end{equation}
Using the commutativity of several of the unitaries in this sequence ($W_{B,e},W_{B,o}$ with anything but $W_E$, $W_M$ with $W_E,W_B$) and defining $W_B=W_{B,o}W_{B,e}$, we obtain the time evolution of a sequence - i.e. a single time step of the digitized evolution:
\begin{equation}
W\left(\tau\right) = W_M W_E W_B W_{oh} W_{ov} W_{eh} W_{ev}
\end{equation}
including only terms which respect the local gauge symmetry, i.e. commute with the $G\left(\mathbf{x}\right)$ operators for every $\mathbf{x}$ separately.

\bibliography{ref}

\end{document}